\documentclass[12pt,epsf]{article}

\usepackage{times}
\usepackage{graphicx}
\usepackage{amsfonts}
\usepackage{amsmath}
\usepackage{amssymb}
\usepackage{amstext}
\usepackage{amsthm}

\usepackage{epsfig}
\usepackage{color}
\psfigdriver{dvips}
\usepackage{latexsym}
\usepackage[matrix,curve]{xypic}
\usepackage{rotating}
\rotdriver{dvips}

\begin{document}

\baselineskip 6mm
\renewcommand{\thefootnote}{\fnsymbol{footnote}}


\newcommand{\nc}{\newcommand}
\newcommand{\rnc}{\renewcommand}


\rnc{\baselinestretch}{1.24}    
\setlength{\jot}{6pt}       
\rnc{\arraystretch}{1.24}   

\makeatletter
\rnc{\theequation}{\thesection.\arabic{equation}}
\@addtoreset{equation}{section}
\makeatother



\nc{\be}{\begin{equation}}

\nc{\ee}{\end{equation}}

\nc{\bea}{\begin{eqnarray}}

\nc{\eea}{\end{eqnarray}}

\nc{\xx}{\nonumber\\}

\nc{\ct}{\cite}

\nc{\la}{\label}

\nc{\eq}[1]{(\ref{#1})}

\nc{\newcaption}[1]{\centerline{\parbox{6in}{\caption{#1}}}}

\nc{\fig}[3]{

\begin{figure}
\centerline{\epsfxsize=#1\epsfbox{#2.eps}}
\newcaption{#3. \label{#2}}
\end{figure}
}


\def\CA{{\cal A}}
\def\CC{{\cal C}}
\def\CD{{\cal D}}
\def\CE{{\cal E}}
\def\CF{{\cal F}}
\def\CG{{\cal G}}
\def\CH{{\cal H}}
\def\CK{{\cal K}}
\def\CL{{\cal L}}
\def\CM{{\cal M}}
\def\CN{{\cal N}}
\def\CO{{\cal O}}
\def\CP{{\cal P}}
\def\CR{{\cal R}}
\def\CS{{\cal S}}
\def\CU{{\cal U}}
\def\CV{{\cal V}}
\def\CW{{\cal W}}
\def\CY{{\cal Y}}
\def\CZ{{\cal Z}}


\def\IB{{\hbox{{\rm I}\kern-.2em\hbox{\rm B}}}}
\def\IC{\,\,{\hbox{{\rm I}\kern-.50em\hbox{\bf C}}}}
\def\ID{{\hbox{{\rm I}\kern-.2em\hbox{\rm D}}}}
\def\IF{{\hbox{{\rm I}\kern-.2em\hbox{\rm F}}}}
\def\IH{{\hbox{{\rm I}\kern-.2em\hbox{\rm H}}}}
\def\IN{{\hbox{{\rm I}\kern-.2em\hbox{\rm N}}}}
\def\IP{{\hbox{{\rm I}\kern-.2em\hbox{\rm P}}}}
\def\IR{{\hbox{{\rm I}\kern-.2em\hbox{\rm R}}}}
\def\IZ{{\hbox{{\rm Z}\kern-.4em\hbox{\rm Z}}}}


\def\a{\alpha}
\def\b{\beta}
\def\d{\delta}
\def\ep{\epsilon}
\def\ga{\gamma}
\def\k{\kappa}
\def\l{\lambda}
\def\s{\sigma}
\def\t{\theta}
\def\w{\omega}
\def\G{\Gamma}


\def\half{\frac{1}{2}}
\def\dint#1#2{\int\limits_{#1}^{#2}}
\def\goto{\rightarrow}
\def\para{\parallel}
\def\brac#1{\langle #1 \rangle}
\def\curl{\nabla\times}
\def\div{\nabla\cdot}
\def\p{\partial}


\def\Tr{{\rm Tr}\,}
\def\det{{\rm det}}


\def\vare{\varepsilon}
\def\zbar{\bar{z}}
\def\wbar{\bar{w}}
\def\what#1{\widehat{#1}}


\def\ad{\dot{a}}
\def\bd{\dot{b}}
\def\cd{\dot{c}}
\def\dd{\dot{d}}
\def\so{SO(4)}
\def\bfr{{\bf R}}
\def\bfc{{\bf C}}
\def\bfz{{\bf Z}}

\begin{titlepage}


\hfill\parbox{3.9cm} {{\tt arXiv:1411.6115}}

\vspace{15mm}

\begin{center}
{\Large \bf Calabi-Yau manifolds from noncommutative \\ Hermitian $U(1)$ instantons}

\vspace{10mm}

Hyun Seok Yang \footnote{hsyang@kias.re.kr}
\\[10mm]

{\sl School of Physics, Korea Institute for Advanced Study,
Seoul 130-722, Korea}

\end{center}

\thispagestyle{empty}

\vskip1cm


\centerline{\bf ABSTRACT}
\vskip 4mm
\noindent

We show that Calabi-Yau manifolds are emergent from the commutative limit of six-dimensional noncommutative
Hermitian $U(1)$ instantons. Therefore, we argue that the noncommutative Hermitian $U(1)$ instantons
correspond to quantized Calabi-Yau manifolds.
\\



\vspace{1cm}

\today

\end{titlepage}

\renewcommand{\thefootnote}{\arabic{footnote}}
\setcounter{footnote}{0}

\section{Introduction to Emergent Gravity}

The emergent gravity can be described by considering the deformation of a symplectic manifold $(M, B)$
where $B$ is a nondegenerate, closed two-form on $M$ \cite{hsy-ijmp09,hsy-jhep09,hsy-jpcs12,q-emg}.
It may be emphasized that a symplectic manifold $(M, B)$ is necessarily an even-dimensional orientable
manifold since $\nu = \frac{1}{n!} B^n$ defines a nowhere vanishing volume form where $\dim(M) = 2n$.
This fact will be important to understand the mirror symmetry of Calabi-Yau (CY) manifolds emergent
from the deformation complex on a symplectic manifold, as will be discussed in a separate
paper \cite{mirror-hsy}. Let us consider a line bundle $L$ over a symplectic manifold $(M, B)$
whose connection one-form is denoted by $A = A_\mu (x) dx^\mu$. The curvature $F$ of a line bundle
is a closed two-form, i.e., $dF=0$ and so locally given by $F = dA$. Suppose that the line
bundle $L$ over $(M, B)$ admits a local gauge symmetry $\mathfrak{B}_L$ which acts on the connection $A$
as well as the symplectic structure $B$ on the base manifold $M$:
\begin{equation}\label{b-tr}
\mathfrak{B}_L : (B, A) \mapsto (B - d\Lambda, A + \Lambda)
\end{equation}
with $\Lambda$ an arbitrary one-form on $M$.
The local gauge symmetry $\mathfrak{B}_L$ is known as the $\Lambda$-symmetry or $B$-field transformation
in string theory. This symmetry then dictates that the curvature $F = dA$ of $L$ appears only
with the combination $\mathcal{F} \equiv B + F$ since the two-form $\mathcal{F}$ is a gauge invariant quantity
under the $\Lambda$-symmetry. Since $d \mathcal{F} = 0$, the line bundle $L$ over $(M, B)$ results
in a ``dynamical" symplectic manifold $(M, \mathcal{F})$
if $\det(1 + F \theta) \neq 0$ where $\theta \equiv B^{-1}$\cite{q-emg}.
Here we mean the ``dynamical" for fluctuating fields around a background.
Therefore the electromagnetic force $F = dA$ manifests itself as the deformation of
a symplectic manifold $(M, B)$.\footnote{\label{2ep}It may be instructive to conceive an analogue
in general relativity. According to the general theory of relativity, the gravitational
force corresponds to the deformation of a given Euclidean space $(M, g)$,
which results in a ``dynamical" Riemannian manifold $(\mathcal{M}, G)$ where $G = g + h$.
The equivalence principle then implies that the ``dynamical" Riemannian manifold $(\mathcal{M}, G)$
can always be trivialized in a locally inertial frame where the metric $G$ recovers the original
unperturbed one $g$. It may be worthwhile to remark that the electromagnetic force is to
the deformation of a symplectic manifold what the gravitational force is to the deformation
of a Riemannian manifold. The emergent gravity picture implies \cite{q-emg} that these
two deformations are isomorphic to each other.}

Since $B$ is a symplectic structure on $M$, it defines a bundle isomorphism
$B: TM \to T^* M$ by $X \mapsto \Lambda = -\iota_X B$ where $X \in \Gamma(TM)$ is
an arbitrary vector field. As a result, the $B$-field transformation (\ref{b-tr}) can be written as
\begin{equation}\label{bx-tr}
\mathfrak{B}_L : (B, A) \mapsto \big( (1+ \mathcal{L}_X) B, A - \iota_X B \big)
\end{equation}
where $\mathcal{L}_X = d \iota_X + \iota_X d$ is the Lie derivative with respect to the vector field $X$.
Note that the ordinary $U(1)$ gauge symmetry, $A \mapsto A + d\lambda$, is a particular case of
the $\Lambda$-symmetry (\ref{b-tr}) for $\Lambda = d\lambda = - \iota_{X_\lambda} B$. In this case,
the vector field $X_\lambda = - \theta(d\lambda)$ is called a Hamiltonian vector field.
Since a vector field is an infinitesimal generator of local coordinate transformations, in other words,
a Lie algebra generator of $\mathrm{Diff}(M)$, the $B$-field transformation (\ref{bx-tr}) can be
identified with a local coordinate transformation generated by the vector field $X \in \Gamma(TM)$.
Consequently the $\Lambda$-symmetry (\ref{b-tr}) can be considered on par with
the (dynamical) diffeomorphism symmetry. This fact leads to a remarkable conclusion \cite{hsy-ijmp09,hsy-jhep09}
that, in the presence of $B$-fields, the underlying local gauge symmetry is rather enhanced.
Thus we fall into a situation similar to general relativity that the dynamical symplectic
manifold $(M, \mathcal{F})$ can be locally trivialized by
a coordinate transformation $\phi \in \mathrm{Diff}(M)$ such that $\phi^* (\mathcal{F}) = B$.
For example, $\phi^* = (1+ \mathcal{L}_X)^{-1} \approx e^{-\mathcal{L}_X}$ if $A = - \Lambda = \iota_X B$.
In other words, it is always possible to find a local coordinate transformation eliminating
dynamical $U(1)$ gauge fields as far as spacetime admits a symplectic structure.
This statement is known as the Darboux theorem or the Moser lemma in symplectic geometry \cite{sg-book}.
It is arguably a novel form of the equivalence principle for the electromagnetic force \cite{q-emg}.
It may be rewarding to revisit the footnote \ref{2ep} with this insight.

Let us introduce an anchor map $\theta = B^{-1}: T^*M \to TM$ defined by $\Lambda
\mapsto X = - \theta(\Lambda)$. The bivector $\theta \in \Gamma(\Lambda^2 TM)$ is called
a Poisson structure on $M$ \cite{sg-book}. It gives the vector space $C^\infty(M)$ a Lie algebra
structure, called a Poisson bracket, which is an antisymmetric, bilinear map
$\{-,-\}_\theta: C^\infty(M) \times C^\infty(M) \to C^\infty(M)$
defined by $(f, g) \mapsto \theta(df, dg) \equiv \{f,g \}_\theta$. An important property is
that the map $f \mapsto  X_f (g) = \{f,g \}_\theta$ is a derivation on $C^\infty(M)$
for any fixed $g \in C^\infty(M)$. In terms of local coordinates on a small patch $U \subset M$,
the Poisson structure $\theta = B^{-1}$ is given by
\begin{equation}\label{poisson-st}
    \theta = \frac{1}{2} \theta^{ab}(y) \frac{\partial}{\partial y^a} \bigwedge
    \frac{\partial}{\partial y^b}.
\end{equation}
Without loss of generality, $\theta^{ab}$ can be chosen to be a constant skew-symmetric matrix
of rank $2n$, typically taking the form $[\theta^{ab}] = \mathbf{1}_n \otimes \sqrt{-1} \theta^i \sigma^2$
with $\theta^i := \theta^{2i-1, 2i}, \; i = 1, \cdots, n$. According to the Darboux theorem stating
that $\phi^* (B+F) = B$, it is always possible to find a {\it locally
inertial frame}, namely, Darboux coordinates, to eliminate the electromagnetic force $F = dA$.
Let us represent the local coordinate transformation $\phi \in \mathrm{Diff}(M)$ as
\begin{equation}\label{darboux-codi}
    \phi: y^a \mapsto x^a(y) = y^a + \theta^{ab} a_b (y).
\end{equation}
The dynamical local coordinates $a_a (y)$ will be called symplectic gauge fields, which are introduced
to compensate local deformations of an underlying symplectic structure by $U(1)$ gauge fields.
The dynamical coordinates $x^a (y)$ are covariant under a symplectic gauge transformation, i.e.,
$\delta x^a (y) = - X_\lambda \big(x^a (y) \big) = \{x^a, \lambda \}_\theta (y)$ and
so play an important role in emergent gravity \cite{hsy-review}.
It is convenient to introduce ``covariant momenta" defined by
\begin{equation}\label{cov-mom}
    D_a (y) \equiv B_{ab} x^b (y) = p_a + a_a (y) \in C^\infty (M)
\end{equation}
where $p_a = B_{ab} y^b$. Note that
\begin{equation}\label{pb-dd}
    \{D_a, D_b \}_\theta = - B_{ab} + f_{ab}
\end{equation}
where $f_{ab} = \partial_a a_b - \partial_b a_a + \{a_a, a_b \}_\theta$ is the field strength
of symplectic gauge fields. One can see that symplectic gauge fields $a_a (y) \in C^\infty (M)$
deform the background Poisson structure specified by $\{p_a, p_b \}_\theta = - B_{ab}$.
In the end, the dynamical symplectic manifold $(M, \mathcal{F})$ can be described by a gauge theory
of symplectic gauge fields introduced via the local coordinate transformation (\ref{darboux-codi}).

Since the symplectic manifold $(M, \mathcal{F})$ is a dynamical system, one may quantize
the system like as quantum mechanics \cite{q-emg}. The quantization is straightforward as the dynamical
system equips with an intrinsic Poisson structure given by (\ref{poisson-st}). An underlying math is
essentially the same as quantum mechanics. It results in a quantized line bundle $\widehat{L}$
over a noncommutative (NC) space \cite{ncft-sw}, denoted by $\mathbb{R}^{2n}_\theta$, 
whose coordinate generators satisfy the commutation relation
\begin{equation}\label{nc-space}
    [y^a, y^b] = i \theta^{ab}.
\end{equation}
The NC $\star$-algebra generated by the Moyal-Heisenberg algebra (\ref{nc-space}) will be denoted
by $\mathcal{A}_\theta$ \cite{ncft-rev}. The quantization $\mathcal{Q}$ also lifts the coordinate
transformation (\ref{darboux-codi}) to a local automorphism of $\mathcal{A}_\theta$ defined
by $\mathcal{Q}: \phi \mapsto \mathcal{D}_A$ which acts on the NC coordinates $y^a$ as \cite{jur-sch}
\begin{equation}\label{dyna-codi}
    \mathcal{D}_A (y^a) \equiv  \widehat{X}^a (y) = y^a + \theta^{ab} \widehat{A}_b (y)
 \in \mathcal{A}_\theta.
\end{equation}
One can see \cite{ncft-rev} that NC $U(1)$ gauge fields are obtained by quantizing symplectic gauge fields,
i.e., $\widehat{A}_a = \mathcal{Q} (a_a)$. Let us define dynamical momentum variables $\widehat{D}_a (y)
\equiv B_{ab} \widehat{X}^b (y) = p_a + \widehat{A}_a (y)$.
Upon quantization, the Poisson bracket is similarly lifted to a NC bracket in $\mathcal{A}_\theta$.
For example, the Poisson bracket relation (\ref{pb-dd}) is now defined by the commutation relation
\begin{equation}\label{ncc-dd}
    -i[\widehat{D}_a, \widehat{D}_b ]_\star = - B_{ab} + \widehat{F}_{ab}
\end{equation}
where the field strength of NC $U(1)$ gauge fields $\widehat{A}_a$ is given by
\begin{equation}\label{nc-curvature}
    \widehat{F}_{ab} = \partial_a \widehat{A}_b - \partial_b \widehat{A}_a
    -i [\widehat{A}_a, \widehat{A}_b]_\star.
\end{equation}
Hence we observe that NC $U(1)$ gauge fields describe a dynamical NC spacetime (\ref{ncc-dd})
which is a deformation of the background NC spacetime (\ref{nc-space}).
To sum up, a dynamical NC spacetime is defined by the quantization of a line bundle $L$ over
a symplectic manifold $(M, B)$ and described by a NC $U(1)$ gauge theory \cite{q-emg}.

An important point \cite{ncft-rev} is that a NC space such as the Moyal-Heisenberg
algebra (\ref{nc-space}) always admits a nontrivial inner automorphism $\mathfrak{A}$
defined by $\mathcal{O} \mapsto \mathcal{O}' = U \star \mathcal{O} \star U^{-1}$
where $U \in \mathfrak{A}$ and $\mathcal{O} \in \mathcal{A}_\theta$.
Its infinitesimal generators consist of an inner derivation $\mathfrak{D}$.
Then there is a well-known Lie algebra homomorphism between the NC $\star$-algebra $\mathcal{A}_\theta$
and the inner derivation $\mathfrak{D}$, defined by the map \cite{hsy-jhep09,hsy-jpcs12,q-emg,hsy-review}
\begin{equation}\label{der-map}
 \mathcal{A}_\theta \to \mathfrak{D}:
 \mathcal{O} \mapsto \mathrm{ad}_{\mathcal{O}} = -i [\mathcal{O}, \cdot \; ]_\star
\end{equation}
for any $\mathcal{O} \in \mathcal{A}_\theta$. Using the Jacobi identity of the NC $\star$-algebra
$\mathcal{A}_\theta$, it is easy to verify the Lie algebra homomorphism:
\begin{equation}\label{lie-homo}
 [ \mathrm{ad}_{\mathcal{O}_1}, \mathrm{ad}_{\mathcal{O}_2} ] =
 -i \mathrm{ad}_{[\mathcal{O}_1, \mathcal{O}_2]_\star}
\end{equation}
for any $\mathcal{O}_1, \mathcal{O}_2 \in \mathcal{A}_\theta$.
In particular, we define the set of NC vector fields given by
\begin{equation}\label{nc-vector}
\{ \widehat{V}_a \equiv \mathrm{ad}_{\widehat{D}_a} \in \mathfrak{D}| \widehat{D}_a
\in \mathcal{A}_\theta, \; a = 1, \cdots, 2n \}
\end{equation}
where $\widehat{D}_a = p_a + \widehat{A}_a$ are previously introduced dynamical NC momenta.
One can apply the Lie algebra homomorphism \eq{lie-homo} to
the commutation relation \eq{ncc-dd} to yield
\begin{equation}\label{homo-f}
\mathrm{ad}_{\widehat{F}_{ab}}  =  [\widehat{V}_a, \widehat{V}_b] \in \mathfrak{D}.
\end{equation}

A basic idea of emergent gravity is to realize the gauge/gravity duality using
the Lie algebra homomorphism (\ref{der-map}) \cite{hsy-jhep09,hsy-jpcs12,q-emg,hsy-review}.
The gauge theory side of the duality is defined by
a NC $U(1)$ gauge theory based on an associative algebra $\mathcal{A}_\theta$ and
its gravity side is defined by associating the derivation $\mathfrak{D}$ of the algebra
$\mathcal{A}_\theta$ with a (quantized) frame bundle of an emergent spacetime manifold $\mathcal{M}$.
But, in order to identify global quantities such as vielbeins in gravity with elements of $\mathfrak{D}$,
it is necessary to glue the local data of the derivation $\mathfrak{D}$ which are derived from
NC gauge fields defined on Darboux charts.
A detailed exposition for the globalization was recently given in Ref. \cite{q-emg}.
See also \cite{jsw-ncl}. For our purpose, it is enough to consider the globalization
for the set of vector fields defined by Eq. (\ref{nc-vector}).
In particular, we are interested in the commutative limit, i.e. $|\theta| \to 0$,
of the derivation algebra $\mathfrak{D}$. In this limit, the NC vector fields in Eq. (\ref{nc-vector})
reduce to ordinary vector fields $V_a = V_a^\mu (y) \partial_\mu \in \Gamma(T\mathcal{M})$, i.e.,
\begin{equation}\label{comm-lim}
    \widehat{V}_a = V_a + \mathcal{O}(\theta^2).
\end{equation}
A $2n$-dimensional manifold emergent from NC $U(1)$ gauge fields will be denoted by $\mathcal{M}$.
The global vector fields $V_a = V_a^\mu (y) \partial_\mu \in \Gamma(T\mathcal{M})$ are related
to inverse vielbeins $E_a = E_a^\mu (y) \partial_\mu \in \Gamma(T\mathcal{M})$
in general relativity as \cite{hsy-jhep09,hsy-jpcs12,q-emg,hsy-review}
\begin{equation}\label{vec-emg}
    V_a = \lambda E_a
\end{equation}
where $\lambda$ is to be determined by a volume-preserving condition. We fix the conformal
factor $\lambda$ by imposing the condition that the vector fields $V_a$ preserve a volume form
\begin{equation}\label{vol-form}
\nu = \lambda^{p} v^1 \wedge \cdots \wedge v^{2n}
\end{equation}
where $v^a = v^a_\mu (y) dy^\mu \in \Gamma(T^* \mathcal{M})$ are coframes dual to $V_a$,
i.e., $\langle v^a, V_b \rangle = \delta^a_b$.
This means that the vector fields $V_a$ obey the condition
\begin{equation}\label{vol-v}
    \mathcal{L}_{V_a} \nu = \big( \nabla \cdot V_a + (p-2n) V_a \ln \lambda \big) \nu = 0.
\end{equation}
Note that any symplectic manifold always admits such volume-preserving vector fields.
See the appendix B in Ref. \cite{q-emg} for the discussion of modular vector fields on a symplectic manifold.
In the end the dynamical metric emergent from NC gauge fields is given by
\begin{equation}\label{emg-metric}
    ds^2 = e^a \otimes e^a = \lambda^2 v^a_\mu (y) v^a_\nu (y) dy^\mu \otimes dy^\nu
\end{equation}
where $e^a = e^a_\mu (y) dy^\mu = \lambda v^a \in \Gamma(T^* \mathcal{M})$ are orthonormal one-forms
on $\mathcal{M}$ and
\begin{equation}\label{vol-lam}
    \lambda^p = \nu (V_1, \cdots, V_{2n}).
\end{equation}

Since the gravitational metric (\ref{emg-metric}) is completely determined by NC $U(1)$ gauge fields,
a spacetime geometry described by the metric (\ref{emg-metric}) will be determined by
the dynamical law of NC $U(1)$ gauge fields. In particular, Einstein gravity emerges
from the commutative limit of NC  $U(1)$ gauge fields \cite{hsy-jhep09,hsy-jpcs12,q-emg,hsy-review}.
This emergent gravity picture may be strengthened
by the fact that a theory of NC $U(1)$ gauge fields respects the diffeomorphism symmetry (\ref{bx-tr})
and the Lie algebra homomorphism (\ref{der-map}) realizes a duality between algebraic objects
in $\mathcal{A}_\theta$ and geometric objects in $\mathfrak{D}$.
For instance, it was recently shown \cite{hea} that a holomorphic line bundle with a nondegenerate
curvature two-form of rank $2n$ is equivalent to a $2n$-dimensional K\"ahler manifold and,
in particular, CY $n$-folds for $n=2$ and $3$ are emergent from the commutative limit of
NC $U(1)$ instantons in four and six dimensions, respectively.
In this paper we will further elaborate the emergent gravity from NC $U(1)$
gauge fields and give a more elegant verification for emergent CY manifolds.

This paper is organized as follows. In Sect. 2, we derive the Hermitian Yang-Mills equations
for NC $U(1)$ gauge fields \cite{non-inst}. In Sect. 3, we show that the commutative limit of
NC Hermitian $U(1)$ instantons is isomorphically mapped to the Hermitian Yang-Mills equations
for spin connections of an emergent CY manifold. In Sect. 4, we argue that the NC
Hermitian $U(1)$ instantons should correspond to quantized CY manifolds which are described
by a matrix model or large $N$ gauge theory \cite{q-emg}. We also briefly discuss the mirror symmetry of
CY manifolds from the emergent gravity picture that will be further illuminated
in a separate paper \cite{mirror-hsy}. In Appendix A, we generalize the emergent gravity picture
in \cite{hea} to the case with world-volume scalar fields as well as $U(1)$ gauge fields.
A main purpose of this appendix is to demystify the emergent gravity picture.
In Appendix B, we present a calculational detail
to verify that the self-duality equations for NC $U(1)$ instantons in four and six dimensions
are transformed to geometrical equations for spin connections of an emergent CY manifold.

\section{NC Hermitian $U(1)$ Instantons}

Let $\widehat{\pi}: \widehat{L} \to \mathbb{R}_\theta^6$ be a NC line bundle over a NC
space $\mathbb{R}_\theta^6$ whose coordinates obey the commutation relation \eq{nc-space}.
Denote the connection of the NC line bundle $\widehat{L}$ by $\widehat{A} = \widehat{A}_a (y) dy^a$
and its curvature $\widehat{F} = \frac{1}{2} \widehat{F}_{ab}  dy^a \wedge dy^b$
is defined by \cite{ncft-sw}
\begin{eqnarray} \label{nc-curv}
\widehat{F} &=& d \widehat{A} - i \widehat{A} \wedge \widehat{A} \nonumber \\
   &=& \frac{1}{2} \big( \partial_a \widehat{A}_b - \partial_b \widehat{A}_a
   - i [\widehat{A}_a, \widehat{A}_b ]_\star \big) dy^a \wedge dy^b.
\end{eqnarray}
The structure group $U(1)_\star$ of $\widehat{L}$ acts on the connection as
\begin{equation}\label{s-group}
 \widehat{A} \mapsto \widehat{A}' =  \widehat{g} \star \widehat{A} \star \widehat{g}^{-1}
 + i \widehat{g} \star d \widehat{g}^{-1}
\end{equation}
where $\widehat{g} \in U(1)_\star$. In order to substantiate the idea that Riemannian manifolds
emerge from NC $U(1)$ gauge fields obeying some equations, let us consider the six-dimensional
NC $U(1)$ gauge theory whose action is given by
\begin{equation}\label{nc-ym}
    S = \frac{1}{4 G^2_{YM}} \int d^6 y \widehat{F}_{ab} \widehat{F}^{ab}.
\end{equation}
We will assume that the multiplication between NC fields is always the star product
if it is not explicitly indicated for a notational simplicity. For example,
$\widehat{f} (y) \widehat{g} (y) := \widehat{f}(y) \star \widehat{g}(y)$ for $\widehat{f},
\widehat{g} \in \mathcal{A}_\theta$.
One can show \ct{cy-hym} that the action \eq{nc-ym} can be written as the Bogomol'nyi form
\begin{eqnarray}\label{bps-cym}
    S &=& \frac{1}{8 G^2_{YM}} \int d^6 y
    \left[\Big( \widehat{F}_{a_1b_1} \pm \frac{1}{4} \varepsilon^{a_1b_1a_2b_2a_3b_3}
  \widehat{F}_{a_2b_2} I_{a_3b_3} \Big)^2 - \frac{1}{2}
  \big(I_{ab} \widehat{F}^{ab} \big)^2 \right. \xx
  && \hspace{3cm} \left. \mp \frac{1}{2} \varepsilon^{a_1b_1a_2b_2a_3b_3}
  \widehat{F}_{a_1b_1} \widehat{F}_{a_2b_2} I_{a_3b_3}
  \right]
\end{eqnarray}
where the constant symplectic matrix $I_{ab}$ is given by
\begin{equation}\label{i-matrix}
    I_{2i-1, 2j} = \delta_{ij} = - I_{2j, 2i-1}, \qquad i,j =1,2,3.
\end{equation}
The above action may be written in a more compact form as
\begin{equation}\label{bps-ym}
    S =  \frac{1}{8 G^2_{YM}} \int d^6 y
     \left[\Big( \widehat{F}_{ab} \pm * (\widehat{F} \wedge B \big)_{ab} \Big)^2
    - \frac{1}{2} \big(I_{ab} \widehat{F}^{ab} \big)^2 \right] \mp \frac{1}{2G^2_{YM}} \int  \widehat{F} \wedge \widehat{F} \wedge B
\end{equation}
where $B = \frac{1}{2}I_{ab} dy^a \wedge dy^b$ is the two-form of rank 6 and will be identified
with the K\"ahler form of $\mathbb{R}^6 \cong \mathbb{C}^3$, i.e., $dB = 0$.
We assume that the wedge product between forms is defined under the star product,
e.g., $\widehat{F} \wedge \widehat{F} = \frac{1}{4} \big( \widehat{F}_{ab}
\star \widehat{F}_{cd} \big) dy^{a} \wedge dy^{b} \wedge dy^{c} \wedge dy^{d}$.
Then, using the Bianchi identity $\widehat{D} \widehat{F} \equiv d \widehat{F}
- i(\widehat{A} \wedge \widehat{F} - \widehat{F} \wedge \widehat{A}) = 0$, one can show that
\begin{equation}\label{top-id}
  \widehat{F} \wedge \widehat{F} = d \widehat{K}
  - \frac{i}{3} (\widehat{C} \wedge \widehat{A} +  \widehat{A} \wedge \widehat{C})
\end{equation}
where $\widehat{K}$ is a NC Chern-Simons term defined by
\begin{equation}\label{boundary-cs}
\widehat{K} \equiv  \widehat{A} \wedge \widehat{F} + \frac{i}{3}
\widehat{A} \wedge \widehat{A} \wedge \widehat{A}
\end{equation}
and
\begin{equation}\label{total-c}
\widehat{C} \equiv  \widehat{A} \wedge \widehat{F} + \widehat{F} \wedge \widehat{A} + \frac{i}{2}
\widehat{A} \wedge \widehat{A} \wedge \widehat{A}.
\end{equation}
Note that the second term in Eq. \eq{top-id} can be written as
\begin{equation}\label{2nd-comm}
- \frac{i}{3} (\widehat{C} \wedge \widehat{A} +  \widehat{A} \wedge \widehat{C}) =
- \frac{i}{3 \cdot 3!} [\widehat{C}_{\mu\nu\rho}, \widehat{A}_\sigma]_\star dy^\mu \wedge dy^\nu
\wedge dy^\rho \wedge dy^\sigma
\end{equation}
and thus it vanishes under the integral thanks to the property \cite{ncft-rev}
\begin{equation}\label{ncint-comm}
    \int d^6 y [\widehat{f}, \widehat{g}]_\star = 0
\end{equation}
for $\widehat{f}, \widehat{g} \in \mathcal{A}_\theta$.
After all, the last term in Eq. (\ref{bps-ym}) can be written as a boundary term
on $\partial \mathbb{R}^6 \cong \mathbb{S}^5$:\footnote{In general, the property \eq{ncint-comm} holds up to
total derivative terms. Hence one may worry that the second term \eq{2nd-comm} may generate an additional
boundary term on $\mathbb{S}^5$. The asymptotic boundary condition for NC $U(1)$ gauge fields is that
$\widehat{F}|_{|y| \to \infty} \to 0$ and so $\widehat{A}|_{|y| \to \infty} \to i \widehat{g} \star
d \widehat{g}^{-1}$ where $\widehat{g} \in U(1)_\star$. Then the topological invariant \eq{boundary-s5}
is coming from the second term in Eq. \eq{boundary-cs} which behaves like $\sim \int_{\mathbb{S}^5}
(\widehat{g} \star d \widehat{g}^{-1})^3 \wedge B$. Note that the commutator term \eq{2nd-comm} contains
more derivatives and thus more rapidly decays at the asymptotic boundary compared to the Chern-Simons
term \eq{boundary-cs}. As a result, total derivative terms in Eq. \eq{2nd-comm} do not contribute
any nontrivial boundary term.}
\begin{equation}\label{boundary-s5}
\mp \frac{1}{2G^2_{YM}} \int_{\mathbb{S}^5} \widehat{K} \wedge B.
\end{equation}

As was shown in Eq. (\ref{boundary-s5}), the last term in Eq. (\ref{bps-ym}) is a topological
term which depends only on the topological class of the vector bundle $\widehat{L}$
over $\mathbb{R}_\theta^6$. Then we can show that the minimum of the action (\ref{bps-ym})
is achieved at the configuration obeying the equations
\begin{equation}\label{nchym1}
    \widehat{F}  = \mp * (\widehat{F} \wedge B)
\end{equation}
or, equivalently, using the fact $*^2 \alpha = \alpha$ for any even form $\alpha$,
\begin{equation}\label{nchym2}
    * \widehat{F}  = \mp \widehat{F} \wedge B.
\end{equation}
Note that the condition $I_{ab} \widehat{F}^{ab} = 0$ needs not be imposed separately because
it can be derived from Eq. \eq{nchym1} by using the identity $\frac{1}{8} {\varepsilon_{ab}}^{cdef}
I_{cd} I_{ef} = I_{ab}$. Therefore the term $I_{ab} \widehat{F}^{ab}$ in Eq. (\ref{bps-ym}) identically
vanishes as long as Eq. (\ref{nchym1}) is satisfied. It may be convenient to write the six-dimensional
version of the self-duality equation (\ref{nchym1}) in terms of component notation:
\begin{equation}\label{nchym3}
    \widehat{F}_{ab}  = \pm \frac{1}{2} {T_{ab}}^{cd} \widehat{F}_{cd}
\end{equation}
where ${T_{ab}}^{cd} \equiv \frac{1}{2} {\varepsilon_{ab}}^{cdef} I_{ef}$.
We will call Eq. (\ref{nchym3}) the Hermitian Yang-Mills (HYM) equations \cite{non-inst}.
It is obvious that a solution of the HYM equations is automatically a solution of the equations of motion,
$\widehat{D}^b \widehat{F}_{ab} = 0$, due to the Bianchi identity, $\widehat{D}_a \widehat{F}_{bc} +
\widehat{D}_b \widehat{F}_{c a} + \widehat{D}_c \widehat{F}_{ab} = 0$.

In six dimensions, the four-form tensor $T_{abcd}$ in Eq. (\ref{nchym3}) breaks
the Lorentz symmetry $SO(6) \cong SU(4)/\mathbb{Z}_2$ to $U(3) = SU(3) \times U(1)$.
Thus it is useful to decompose the 15-dimensional vector space of two-forms $\Lambda^2 T^*\mathcal{M}$ under
the unbroken symmetry group $U(3)$ into three subspaces \cite{cy-hym}:\footnote{This decomposition
can easily be understood by the Clifford isomorphism $\mathbb{C}l(d) = \bigoplus_{k=0}^d \mathbb{C}l^k (d)
\cong \Lambda^* \mathcal{M} = \bigoplus_{k=0}^d \Lambda^k T^* \mathcal{M}$,
stating that there exists a vector space isomorphism between the Clifford algebra $\mathbb{C}l(d)$
in $d$-dimensions and the exterior algebra $\Lambda^* \mathcal{M}$ of cotangent bundle $T^*\mathcal{M}$
over $\mathcal{M}$. In particular, the Clifford isomorphism implies that the Lorentz generators $J^{ab}
= \frac{1}{4} [\Gamma^a, \Gamma^b]$ in $\mathbb{C}l(d)$ are in one-to-one correspondence with two-forms in
the vector space $\Lambda^2 T^* \mathcal{M}$. Then the decomposition (\ref{dec23}) is simply the branching
of the vector space $\Lambda^2 T^* \mathcal{M}$ under the symmetry reduction $SO(6) \to U(3) = SU(3) \times U(1)$.
It was argued in \cite{cy-hym} that the Clifford isomorphism leads to an elegant picture for the mirror symmetry
of CY manifolds.}
\begin{equation}\label{dec23}
 \Lambda^2 T^*\mathcal{M} = \Lambda^2_1 \oplus \Lambda^2_6 \oplus \Lambda^2_8
\end{equation}
where $\Lambda^2_1, \; \Lambda^2_6$, and $\Lambda^2_8$ are one-dimensional (singlet), six-dimensional and
eight-dimensional vector spaces taking values in $U(1) \subset U(3), \; SU(4)/U(3)
= \mathbb{C}\mathbb{P}^3$, and $SU(3) \subset U(3)$, respectively.
One can show \cite{cy-hym} that, under the choice \eq{i-matrix} for the K\"ahler form $B$ of $\mathbb{R}^6$,
the NC $U(1)$ gauge fields in $\Lambda^2_6$, i.e. $\widehat{F} = \frac{1}{2} \widehat{F}_{ab}
dy^a \wedge dy^b \in \Lambda^2_6$, obey the $(+)$-equations
\begin{equation}\label{nchym+}
    \widehat{F}_{ab}  = \frac{1}{2} {T_{ab}}^{cd} \widehat{F}_{cd}
\end{equation}
while, if $\widehat{F} \in \Lambda^2_8$, they obey the $(-)$-equations
\begin{equation}\label{nchym-}
    \widehat{F}_{ab}  = - \frac{1}{2} {T_{ab}}^{cd} \widehat{F}_{cd}.
\end{equation}
Explicitly, the $(-)$-equations \eq{nchym-}, for example, are given by
\begin{eqnarray} \label{comp-holo}
  &&  \widehat{F}_{2i-1, 2j-1} =  \widehat{F}_{2i, 2j}, \qquad     \widehat{F}_{2i-1, 2j} =
 -  \widehat{F}_{2i, 2j-1}, \qquad i, j = 1, 2, 3, \\
 \label{comp-stab}
  && \frac{1}{2} I^{ab} \widehat{F}_{ab}
  = \widehat{F}_{12} + \widehat{F}_{34} + \widehat{F}_{56} = 0.
\end{eqnarray}
But note that, for the case of the $(+)$-equations \eq{nchym+}, Eq. (\ref{comp-holo}) has sign flips
and Eq. (\ref{comp-stab}) is replaced by $\widehat{F}_{12} = \widehat{F}_{34} = \widehat{F}_{56} = 0$.
Hence it consists of totally nine equations and so only six components in $\Lambda^2_6$ remain.
For the reason to be explained later, NC Hermitian $U(1)$ instantons or shortly NC instantons are given
by the solutions of the $(-)$-equation \eq{nchym-} only.
After all, the NC Hermitian $U(1)$ instantons are constructed by projecting the vector
space $\Lambda^2 T^* \mathcal{M}$ into the eight-dimensional subspace $\Lambda^2_8$ which
respects the $SU(3)$ rotational symmetry. Hence it may be naturally expected that CY manifolds with
$SU(3)$ holonomy emerge from the NC Hermitian $U(1)$ instantons obeying the $(-)$-equation \eq{nchym-}
rather than the $(+)$-equation \eq{nchym+}.

The symplectic structure (\ref{i-matrix}) provides a natural pairing between coordinates which
picks up a particular complex structure on $\mathbb{R}^6$.
The complex coordinates specified by the symplectic matrix (\ref{i-matrix}) are given by\footnote{From
now on, the imaginary unit $i$ will be denoted by $\sqrt{-1}$ to avoid a confusion
with the frequently appearing holomorphic index $i$.}
\begin{equation}\label{comp-coord}
    z^i = y^{2i-1} + \sqrt{-1} y^{2i}, \qquad \overline{z}^{\bar{i}} = y^{2i-1} - \sqrt{-1} y^{2i}, \qquad
    i, \bar{i} = 1, 2, 3
\end{equation}
and the corresponding NC $U(1)$ gauge fields take the combination
\begin{equation}\label{comp-u1g}
    \widehat{A}_i = \frac{1}{2} \bigl( \widehat{A}_{2i-1} - \sqrt{-1} \widehat{A}_{2i} \bigr),
    \qquad \widehat{A}_{\overline{i}}
    = \frac{1}{2} \bigl( \widehat{A}_{2i-1} + \sqrt{-1} \widehat{A}_{2i} \bigr).
\end{equation}
Then the field strengths of $(2,0)$ and $(1,1)$ parts are, respectively, given by
\begin{eqnarray} \label{fs-20}
&& \widehat{F}_{ij} = \frac{1}{4} \bigl( \widehat{F}_{2i-1, 2j-1} -  \widehat{F}_{2i, 2j} \bigr)
- \frac{\sqrt{-1}}{4} \bigl( \widehat{F}_{2i-1, 2j} +  \widehat{F}_{2i, 2j-1} \bigr), \\
\label{fs-11}
&& \widehat{F}_{i\overline{j}} = \frac{1}{4} \bigl( \widehat{F}_{2i-1, 2j-1} +  \widehat{F}_{2i, 2j} \bigr)
+ \frac{\sqrt{-1}}{4} \bigl( \widehat{F}_{2i-1, 2j} -  \widehat{F}_{2i, 2j-1} \bigr).
\end{eqnarray}
Therefore Eq. (\ref{comp-holo}) can be written as
\begin{equation}\label{holo-vb}
\widehat{F}_{ij} = \widehat{F}_{\overline{i}\overline{j}} = 0
\end{equation}
which means that NC $U(1)$ gauge fields obeying the HYM equations (\ref{nchym-})
must be a connection of a NC holomorphic line bundle.
The last equation (\ref{comp-stab}) is equivalent to the condition
\begin{equation}\label{stab-vb}
\sum_{i=1}^3 \widehat{F}_{i \overline{i}} = 0
\end{equation}
which corresponds to the stability of the NC holomorphic line bundle \cite{non-inst}.

For the $(+)$-equation \eq{nchym+}, we get instead
\begin{equation}\label{+-hym}
\widehat{F}_{i \overline{j}} = 0
\end{equation}
and the totally six components from $ \widehat{F}_{ij}$ and $\widehat{F}_{\overline{i}\overline{j}}$
survive. Therefore the NC $U(1)$ gauge fields obeying the $(+)$-equation \eq{nchym+} will not give rise
to a NC Hermitian $U(1)$ instanton.

\section{Emergent Calabi-Yau Manifolds}

Using the Lie algebra homomorphism \eq{homo-f}, one can translate the HYM equations \eq{nchym3}
into some geometric equations for the vector fields determined by NC $U(1)$ gauge fields
in Eq. \eq{nc-vector} \cite{hsy-jhep09,hsy-jpcs12,q-emg,hsy-review}.
For instance, the $(-)$-equations \eq{nchym-} are equivalently stated as
\begin{equation}\label{nvbps}
    [\widehat{V}_a, \widehat{V}_b] = - \frac{1}{2} {T_{ab}}^{cd} [\widehat{V}_c, \widehat{V}_d].
\end{equation}
In the commutative limit \eq{comm-lim}, NC vector fields $\widehat{V}_a$ reduce to ordinary vector fields
and Eq. \eq{nvbps} in this limit is defined by the Lie bracket. Hence we introduce the Lie bracket
of the vector fields $V_a \in \Gamma(T \mathcal{M})$ defined by
\begin{equation}\label{cvfse}
    [V_a, V_b] = - {g_{ab}}^{c} V_c.
\end{equation}
The Lie algebra \eq{cvfse} means that the dual covectors $v^a \in \Gamma(T^* \mathcal{M})$ obey the structure
equations $d v^a = \frac{1}{2} {g_{bc}}^{a} v^b \wedge v^c$.
Using Eq. \eq{cvfse}, the commutative limit of Eq. \eq{nvbps} can be written as
\begin{equation}\label{csebps}
    {g_{ab}}^{e} = - \frac{1}{2} {T_{ab}}^{cd} {g_{cd}}^{e}.
\end{equation}
In Appendix B, we show that, for a particular choice of volume form \eq{vol-form},
Eq. \eq{csebps} can be transformed into simple equations for spin connections given by
\begin{equation}\label{bps-spin}
    \omega_{ab} = - \frac{1}{2} {T_{ab}}^{cd} \omega_{cd}.
\end{equation}
Now we will show that a six-dimensional manifold obeying Eq. \eq{bps-spin} must be
a Ricci-flat, K\"ahler manifold which is called a CY manifold \cite{gsw-string,besse}.
Therefore we come to a conclusion that CY manifolds are emergent from the commutative limit
of six-dimensional NC Hermitian $U(1)$ instantons.

Suppose that $\mathcal{M}$ is a six-dimensional complex manifold whose metric $ds^2 = g_{\mu\nu} dy^\mu
\otimes dy^\nu$ is given by Eq. (\ref{emg-metric}). Let us choose local complex coordinates $y^\mu = (z^\alpha, \overline{z}^{\bar{\alpha}}), \; \alpha = 1, 2, 3$ in which an almost complex structure takes
the form ${J^\alpha}_\beta = \sqrt{-1} {\delta^\alpha}_\beta, \; {J^{\bar{\alpha}}}_{\bar{\beta}}
= - \sqrt{-1} {\delta^{\bar{\alpha}}}_{\bar{\beta}}$. This complex structure is basically inherited from
the symplectic structure $J = I = \mathbf{1}_3 \otimes \sqrt{-1} \sigma^2$ which we already introduced
in Eq. (\ref{i-matrix}). We have split a curved space index $\mu = 1, \cdots, 6 = (\alpha, \bar{\alpha})$
into a holomorphic index $\alpha = 1, 2, 3$ and an anti-holomorphic one $\bar{\alpha} = 1, 2, 3$ and
similarly, a tangent space index $a = 1, \cdots, 6 = (i, \bar{i})$ into $i = 1, 2, 3$ and $\bar{i} = 1, 2, 3$.
We impose the Hermitian condition on the complex manifold $\mathcal{M}$ defined by $g(X, Y) = g(JX, JY)$
for any $X, Y \in \Gamma(T\mathcal{M})$ \cite{besse}.
This means that the metric $g$ on the complex manifold $\mathcal{M}$
is a Hermitian metric, i.e., $g_{\alpha\beta} = g_{\bar{\alpha}\bar{\beta}} = 0, \;
g_{\alpha\bar{\beta}} = g_{\bar{\beta}\alpha}$. The Hermitian condition can be solved
by taking the vielbeins as
\begin{equation}\label{hermitian-viel}
    e^i_{\bar{\alpha}} = e^{\bar{i}}_\alpha = 0 \quad \mathrm{and}  \quad
    E_i^{\bar{\alpha}} = E_{\bar{i}}^\alpha = 0.
\end{equation}
Then the two-form defined by $\Omega = \frac{1}{2} I_{ab} e^a \wedge e^b$ is a K\"ahler form, i.e.,
$\Omega(X, Y) = g(X, JY)$ and it is given by
\begin{equation}\label{kahler-form}
    \Omega = - \sqrt{-1} e^i \wedge e^{\bar{i}} = - \sqrt{-1} e^i_{\alpha} e^{\bar{i}}_{\bar{\beta}}
    dz^\alpha \wedge d\overline{z}^{\bar{\beta}}
    = - \sqrt{-1} g_{\alpha\bar{\beta}}  dz^\alpha \wedge d\overline{z}^{\bar{\beta}}.
\end{equation}
Using the torsion free condition, $de^a + {\omega^a}_b \wedge e^b = 0$, it is easy to show \cite{gsw-string}
that the K\"ahler condition, $d\Omega = 0$, is equivalent to the one
that the spin connection $\omega_{ab}$ on a Hermitian manifold $(\mathcal{M}, g)$ is $U(3)$-valued, i.e.,
\begin{equation}\label{u3-spin}
    \omega_{ij} = \omega_{\bar{i}\bar{j}} = 0.
\end{equation}
In this case, the K\"ahler metric is solely determined by a K\"ahler potential $K(z, \overline{z})$ as
\begin{equation}\label{kahler-metric}
    g_{\alpha\bar{\beta}} = \partial_\alpha \overline{\partial}_{\bar{\beta}}
    K(z, \overline{z}).
\end{equation}

It is well-known \cite{gsw-string,besse} that the Ricci tensor of a K\"ahler manifold is the field strength
of the $U(1)$ part of $U(3) = SU(3) \times U(1)$ spin connections and the $U(1)$ gauge field is given
by the trace part of $U(3)$ spin connections, i.e.,
\begin{equation}\label{spin-u1}
    A^{(0)} \equiv \sqrt{-1} \sum_{i=1}^3 \omega_{\bar{i}i}.
\end{equation}
Therefore a K\"ahler manifold $(\mathcal{M}, g)$ is Ricci-flat if $F^{(0)} = d A^{(0)} = 0$
or $A^{(0)} = d \lambda$. Note that the first cohomology for a simply connected manifold $\mathcal{M}$
identically vanishes, i.e., $H^1(\mathcal{M}) = 0$. Hence it is possible to choose a gauge, $A^{(0)} = 0$,
for a simply connected Ricci-flat manifold which means that
\begin{equation}\label{u1=0}
    \sum_{i=1}^3 \omega_{\bar{i}i} = 0.
\end{equation}
A Ricci-flat and K\"ahler manifold is known as a CY manifold which plays an important role in string theory compactification \cite{gsw-string}.
Consequently a CY manifold with $SU(3)$ holonomy is characterized, up to a gauge choice,
by Eqs. \eq{u3-spin} and \eq{u1=0}. In terms of real coordinates, they are succinctly summarized
by Eq. \eq{bps-spin}. In Appendix B, we show that the (generalized) self-duality equation \eq{bps-spin}
for spin connections is equivalent to the commutative limit of NC Hermitian $U(1)$ instantons defined by
the Hermitian Yang-Mills equation \eq{nchym-} in six dimensions.
Therefore we see that the commutative limit of six-dimensional NC Hermitian $U(1)$ instantons
can be rephrased into the Ricci-flat and K\"ahler condition for Calabi-Yau manifolds.

\section{Discussion}

Suppose that $\mathcal{F} = d \mathcal{A} = B + F$ is the curvature of a holomorphic line bundle, i.e.,
$\mathcal{F}_{ij} = \partial_i \mathcal{A}_j - \partial_j \mathcal{A}_i = 0$ and
$\mathcal{F}_{\bar{i}\bar{j}} = \overline{\partial}_{\bar{i}} \mathcal{A}_{\overline{j}}
- \overline{\partial}_{\bar{j}} \mathcal{A}_{\overline{i}} = 0$. It can be solved by
$\mathcal{A}_i = \frac{\sqrt{-1}}{2} \partial_i \phi(z, \overline{z})$ and
$\mathcal{A}_{\overline{i}} = - \frac{\sqrt{-1}}{2} \overline{\partial}_{\bar{i}} \phi(z, \overline{z})$
where $\phi(z, \overline{z})$ is a real smooth function on $\mathbb{C}^n$.
Then the curvature of a holomorphic line bundle is given by
\begin{equation}\label{holo-curv}
 \mathcal{F} = - \sqrt{-1} \partial_i \overline{\partial}_{\bar{j}}  \phi(z, \overline{z})
 dz^i \wedge d\overline{z}^{\bar{j}} = - \sqrt{-1} \partial \overline{\partial} \phi(z, \overline{z}).
\end{equation}
Note that the K\"ahler form \eq{kahler-form} of a $2n$-dimensional K\"ahler manifold is given by
\begin{equation}\label{2n-kahler}
 \Omega =  - \sqrt{-1} \partial \overline{\partial} K(z, \overline{z}).
\end{equation}
It was shown in the Appendix of \cite{hea} that one can identify $\phi(z, \overline{z})$
and $K(z, \overline{z})$ if the curvature $\mathcal{F}$ of a holomorphic line bundle
is a symplectic structure, i.e.,
a nondegenerate, closed two-form. This means that a holomorphic line bundle with a nondegenerate
curvature two-form of rank $2n$ is equivalent to a $2n$-dimensional K\"ahler manifold.
A CY manifold is a K\"ahler manifold with a vanishing first Chern class \cite{besse}.
In this paper we have verified a particular case for the equivalence between a K\"ahler manifold
and a holomorphic line bundle.

Since we have considered a line bundle over $\mathbb{R}^6$ with a symplectic structure $B$,
the emergent CY manifolds from NC $U(1)$ gauge fields are noncompact.
But the result can be generalized to compact CY manifolds by considering a line bundle $L \to M$
over a compact K\"ahler manifold $M$ with a symplectic structure $B$ if the two-form $B$ is
a K\"ahler form of the base manifold $M$, although an explicit construction may be more difficult.
Actually the argument in \cite{hea} can be generalized to a holomorphic line bundle over
a compact K\"ahler manifold $M$ whose K\"ahler structure is given by a background $B$-field.
Then the result in \cite{hea} will be equally applied to the compact case.

We observed in Sect. 1 that a line bundle over a symplectic manifold $(M, B)$ results in
a {\it dynamical} symplectic manifold $(M, \mathcal{F})$. The quantization of the dynamical
symplectic manifold $(M, \mathcal{F})$ gives rise to a dynamical NC spacetime described by
a NC $U(1)$ gauge theory \cite{q-emg}. Therefore NC $U(1)$ gauge fields correspond to a {\it quantized}
dynamical spacetime. In this paper we showed that CY manifolds are emergent from a semi-classical limit
of NC Hermitian $U(1)$ instantons. Note that the ordinary vector field $V_a \in \Gamma(T\mathcal{M})$
in Eq. \eq{comm-lim} is just the leading part of the NC vector field $\widehat{V}_a \in
\mathfrak{D}$ when the commutative limit is taken into account and a classical Riemannian manifold
$\mathcal{M}$ is constructed by the set of the usual vector fields $V_a$. Thus it is natural to think
of NC Hermitian $U(1)$ instantons as a quantized geometry of CY manifolds.
Since NC gauge fields can be represented by large $N$ matrices in $\mathrm{End} (\mathcal{H})$
where $\mathcal{H}$ is a Hilbert space representing the NC space \eq{nc-space} \cite{ncft-rev,hsy-epjc09},
the quantized CY manifold will be described by a matrix theory or large $N$ gauge theory.
It is well-known \cite{m-model,ads-cft} that such a matrix model or large $N$ gauge theory describes
a nonperturbative formulation of string/M theories.
Therefore it is reasonable \cite{q-emg} that NC $U(1)$ gauge fields describe a quantum geometry.

We explained in Sect. 1 that emergent gravity is defined by considering the deformation of
a symplectic manifold $(M, B)$ by a line bundle $L \to M$. The line bundle $L$ manifests itself
only by introducing a new symplectic structure $\mathcal{F} = B + F$ where $F = dA$ is identified with
the curvature of the line bundle \cite{q-emg}. Then symplectic or NC $U(1)$ gauge fields are introduced via
a local coordinate transformation $\phi \in \mathrm{Diff}(M)$ eliminating dynamical $U(1)$ gauge fields,
i.e., $\phi^* (\mathcal{F}) = B$. The underlying math for this argument is the well-known theorem
in symplectic geometry known as the Darboux theorem \cite{sg-book}.
Note that a CY manifold $X$ always arises with a mirror pair $Y$ obeying the mirror relation \cite{gsw-string}
\begin{equation}\label{mirror}
    h^{1,1} (X) = h^{2,1} (Y), \qquad h^{2,1} (X) = h^{1,1} (Y)
\end{equation}
where $h^{p,q}$ is a Hodge number of a CY manifold.
Since we showed that six-dimensional CY manifolds are emergent from
the commutative limit of NC Hermitian $U(1)$ instantons which are the connections in a stable
holomorphic line bundle $L \to M$ \cite{non-inst}, an interesting question arises when we conceive
the emergent CY manifolds from the mirror symmetry perspective.
What is the mirror symmetry from the emergent gravity picture ?
Emergent gravity seems to provide a very elegant picture for the mirror symmetry \cite{cy-hym}.
Note that a symplectic manifold $(M, B)$ is necessarily an orientable manifold and so admits
the Hodge dual operation $*: \Omega^k (M) \to \Omega^{6-k} (M)$ between vector spaces
of $k$-forms and $(6-k)$-forms. Suppose that $C$ is a nondegenerate four-form that is co-closed,
i.e., $\delta C = 0$ where $\delta = - * d *: \Omega^k (M) \to \Omega^{k-1} (M)$ is
the adjoint exterior differential operator. Define a two-form
$\widetilde{B} \equiv * C$. Then $\delta C = 0 \Leftrightarrow d\widetilde{B} = 0$.
Therefore $\widetilde{B}$ defines another symplectic structure independent of $B$.
Hence one can equally consider the deformation of the {\it dual} symplectic structure $\widetilde{B}$
by considering a dual line bundle $\widetilde{L} \to M$. The curvature $\widetilde{F} = d \widetilde{A}$
of the dual line bundle $\widetilde{L}$ may be identified with the Hodge dual of
a co-closed four-form $G$, i.e., $\widetilde{F} = *G$. Then we have the property $\delta G = 0
\Leftrightarrow d\widetilde{F} = 0$. Therefore the dual line bundle $\widetilde{L}$ similarly results in
a dynamical symplectic manifold $(M, \widetilde{\mathcal{F}})$ where $\widetilde{\mathcal{F}}
= \widetilde{B} + \widetilde{F} = *(C+G)$. One can introduce {\it dual} NC $U(1)$ gauge fields by
a local coordinate transformation $\widetilde{\phi} \in \mathrm{Diff}(M)$ such that $\widetilde{\phi}^*
(\widetilde{\mathcal{F}}) = \widetilde{B}$. After all it should be possible to find {\it dual}
NC Hermitian $U(1)$ instantons as a solution of Hermitian Yang-Mills equations
defined by dual NC $U(1)$ gauge fields. It is obvious that a CY manifold will also arise from
the dual NC Hermitian $U(1)$ instanton which is independent of a CY manifold emergent from the line bundle
$L$ over the symplectic manifold $(M, B)$. In other words, the variety of emergent CY manifolds
is doubled thanks to the Hodge duality $*: \Omega^4 (M) \to \Omega^{2} (M)$. Since two classes of emergent
CY manifolds are independent of each other, it should be possible
to arrange a pair $(X, Y)$ such that $\chi (X) = - \chi(Y)$ where $\chi(M)$ is the Euler characteristic
of a CY manifold $M$. Since $\chi(M) = 2 \big(h^{1,1} (M) - h^{2,1} (M) \big)$ and $h^{p, q} (M) \geq 0$,
$\chi (X) = - \chi(Y)$ implies the mirror relation \eq{mirror}. Consequently, the emergent gravity
suggests a beautiful picture that the mirror symmetry of CY manifolds is simply the Hodge theory for
the deformation of symplectic and dual symplectic structures.
We will discuss the mirror symmetry in emergent gravity elsewhere \cite{mirror-hsy}.

\section*{Acknowledgments}

We thank Sangheon Yun for helpful discussions and collaboration at an early stage
of this work. This work was supported by the National Research Foundation of Korea (NRF) grant
funded by the Korea government (MOE) (No. 2011-0010597).


\appendix

\section{Emergent gravity demystified}

A main purpose of this appendix is to demystify the emergent gravity picture.
We will generalize the emergent gravity picture to the case with world-volume scalar fields as well as
$U(1)$ gauge fields. The underlying argumentation is a straightforward generalization of the method
in Ref. \cite{hea}. For simplicity, we will consider the four-dimensional Dirac-Born-Infeld (DBI) action
given by
\begin{equation}\label{s-dbi}
    S = \frac{1}{2\pi \kappa^2 g_s} \int d^4 x \sqrt{\det \big(h + \kappa(B+F) \big)}
    + \mathcal{O} (\sqrt{\kappa} \partial F, \cdots),
\end{equation}
where $\kappa = 2\pi \alpha'$ and $h_{\mu\nu} = g_{\mu\nu} + \kappa^2 \partial_\mu \phi^a
\partial_\nu \phi^a$ is a world-volume metric in static gauge. We will not assume that the metric
$g_{\mu\nu}$ is flat. Note that the worldvolume scalar fields
$\phi^a \; (a=1, \cdots, n)$ carry the mass dimension. The four-dimensional result can be easily generalized
to higher dimensions.

In a low energy limit where $\kappa \to 0$, using the formula
\begin{equation}\label{det-formu}
\sqrt{\det (1+A)} \approx 1
+ \frac{1}{2} \mathrm{Tr} A - \frac{1}{4} \mathrm{Tr} A^2 + \frac{1}{8} (\mathrm{Tr} A)^2 + \cdots,
\end{equation}
where $A_{\mu\nu} = \kappa(B_{\mu\nu}+F_{\mu\nu}) + \kappa^2 \partial_\mu \phi^a \partial_\nu \phi^a$,
we can expand the DBI action \eq{s-dbi} in powers of $\kappa$:
\begin{equation}\label{es-dbi}
    S = \frac{V_4}{2\pi \kappa^2 g_s} + \frac{1}{2\pi g_s} \int d^4 x \sqrt{g} \Big( \frac{1}{2}
    \partial_\mu \phi^a \partial^\mu \phi^a + \frac{1}{4} (F_{\mu\nu} + B_{\mu\nu}) (F^{\mu\nu} + B^{\mu\nu})
    + \mathcal{O} (\kappa) \Big),
\end{equation}
where $V_4$ is a world-volume of a D3-brane and we suppressed higher-order terms. Therefore
the DBI action \eq{s-dbi} describes a $U(1)$ gauge theory with neutral scalar fields in the background $B$ field.
However we may form complex scalar fields by $\psi_1 = \phi_1 + i \phi_2, \; \psi_2 = \phi_3 + i \phi_4$, etc.
Using the Darboux theorem in symplectic geometry \cite{sg-book},
we can always find local coordinates $\phi: y \mapsto x=x(y)$ such that
\begin{equation}\label{app-darboux}
    \big( B_{\alpha\beta}+F_{\alpha\beta} (x) \big ) \frac{\partial x^\alpha}{\partial y^\mu}
    \frac{\partial x^\beta}{\partial y^\nu} = B_{\mu\nu}.
\end{equation}
This well-known theorem leads to a remarkable identity:
\begin{eqnarray}\label{sdbi-id}
    S &=& \frac{1}{2\pi \kappa^2 g_s} \int d^4 x \sqrt{\det \big(h + \kappa(B+F) \big)} \nonumber\\
    &=& \frac{1}{2\pi \kappa^2 g_s} \int d^4 y \sqrt{\det \big(\mathcal{G} + \kappa B \big)},
\end{eqnarray}
where $\mathcal{G}_{\mu\nu}$ is a dynamical metric given by
\begin{equation}\label{d-metric}
  \mathcal{G}_{\mu\nu} =  g_{\alpha\beta}\frac{\partial x^\alpha}{\partial y^\mu}
  \frac{\partial x^\beta}{\partial y^\nu}
  + \kappa^2 \frac{\partial \phi^a}{\partial y^\mu} \frac{\partial \phi^a}{\partial y^\nu}.
\end{equation}
It should be remarked that the coordinate transformation \eq{app-darboux} to a Darboux frame is locally defined
and accordingly the dynamical metric \eq{d-metric} is also locally defined. However we can glue
the local coordinate patches together to yield a globally defined Riemannian metric.
See Ref. \cite{q-emg} for a detailed exposition on the globalization of emergent geometry.
We will assume such a globalization for our local constructions.

The metric \eq{d-metric} suggests that it will be convenient to introduce the embedding coordinates
defined by
\begin{equation}\label{embed-coord}
X^M = (x^\alpha, \kappa \phi^a), \qquad M = 1, \cdots, n+4,
\end{equation}
and rewrite the metric as the form
\begin{equation}\label{exd-metric}
  \mathcal{G}_{\mu\nu} =  G_{MN} \frac{\partial X^M}{\partial y^\mu} \frac{\partial X^N}{\partial y^\nu}
\end{equation}
with $G_{MN} = (g_{\alpha\beta}, \delta_{ab})$.
Since the coordinate transformations in Eq. \eq{app-darboux} are eliminating dynamical $U(1)$ gauge fields,
the coordinates $x^\alpha = x^\alpha(y)$ must be {\it dynamical} which may be represented by
\begin{equation}\label{dyn-coor}
    x^\alpha(y) = \theta^{\alpha\beta} \big(p_\beta + \widehat{A}_\beta (y) \big)
\end{equation}
where $\theta \equiv B^{-1}$ and $p_\alpha = B_{\alpha\beta} y^\beta$.
Now we will show that the DBI action \eq{sdbi-id} can be written as the action of NC $U(1)$ gauge fields and
adjoint scalar fields. We want to expand the second action around the background $B$ field, i.e.,
\begin{equation}\label{expan1}
\sqrt{\det \big(\mathcal{G} + \kappa B \big)} = \sqrt{\det (\kappa B)}
\sqrt{\det \big(1 + A \big)},
\end{equation}
where ${A_\mu}^\nu = \frac{1}{\kappa} \mathcal{G}_{\mu\lambda} \theta^{\lambda\nu}$.
Since $\sqrt{\det \big(\mathcal{G} + \kappa B \big)} = \sqrt{\det \big(\mathcal{G} - \kappa B \big)}$,
the even powers of $\kappa$ will only contribute to the expansion.
Using the formula $\sqrt{\det \big(1 + A \big)} = \exp \sum_{k=1}^\infty \frac{(-)^{k+1}}{2k} \mathrm{Tr} A^k$,
it is straightforward to show that
\begin{equation}\label{expan2}
\sqrt{\det_4 \big(1 + A \big)} = \sqrt{\det_{n+4} \Big(1 + \frac{1}{\kappa} G \mathfrak{P} \Big)},
\end{equation}
where the subscript in the determinant indicates the size of matrix and
we introduced the Poisson bracket defined by
\begin{equation}\label{app-poisson}
    \mathfrak{P}^{MN} = \{ X^M, X^N \}_\theta \equiv \theta^{\mu\nu}
    \frac{\partial X^M}{\partial y^\mu} \frac{\partial X^N}{\partial y^\nu}.
\end{equation}
Using the expression \eq{dyn-coor}, we can explicitly calculate the above Poisson bracket to yield
\begin{eqnarray}\label{exp-poissonbr}
    \mathfrak{P}^{MN} &=& \left(
                          \begin{array}{cc}
    \big( \theta(B-\widehat{F}) \theta \big)^{\mu\nu} & \kappa \theta^{\mu\nu} \widehat{D}_\nu \phi^b \\
                            \kappa \widehat{D}_\mu \phi^a \theta^{\mu\nu} & \kappa^2 \{ \phi^a, \phi^b \}_\theta \\
                          \end{array}
                        \right) \nonumber \\
                        &=& \left(
                                    \begin{array}{cc}
                                      - \theta^{\mu\rho} & 0 \\
                                      0 & \kappa \delta^{ac} \\
                                    \end{array}
                                  \right)
                                  \left(
                          \begin{array}{cc}
                            (\widehat{F} - B)_{\rho\sigma} & - \widehat{D}_\rho \phi^d \\
                            \widehat{D}_\sigma \phi^c & \{ \phi^c, \phi^d \}_\theta \\
                          \end{array}
                        \right)
                              \left(
                                    \begin{array}{cc}
                                      \theta^{\sigma\nu} & 0 \\
                                      0 & \kappa \delta^{db} \\
                                    \end{array}
                                  \right),
\end{eqnarray}
where the covariant derivative and the field strength of symplectic gauge fields in Eq. \eq{dyn-coor}
are given by $\widehat{D}_\mu \phi^a = \partial_\mu \phi^a + \{\widehat{A}_\mu, \phi^a \}_\theta$
and $\widehat{F}_{\mu\nu} = \partial_\mu \widehat{A}_\nu - \partial_\nu \widehat{A}_\mu
+ \{\widehat{A}_\mu, \widehat{A}_\nu \}_\theta$, respectively.

Combining the above results together, Eq. \eq{expan1} can be arranged into the form
\begin{eqnarray} \label{app-id1}
   \sqrt{\det_4 \big(\mathcal{G} + \kappa B \big)} &=& \sqrt{\frac{\det_4 (\kappa B)}{\det_4 \mathfrak{g}}}
   \sqrt{\det_{n+4} \big(\mathbb{G} + \kappa \mathfrak{G} \big)} \nonumber \\
  &=& \frac{g_s}{G_s} \sqrt{\det_{n+4} \big(\mathbb{G} + \kappa \mathfrak{G} \big)}
\end{eqnarray}
where
\begin{equation}\label{2matrix}
    \mathbb{G}_{MN} = \left(
                   \begin{array}{cc}
                     \mathfrak{g}_{\mu\nu} & 0 \\
                     0 & \delta_{ab} \\
                   \end{array}
                 \right), \qquad \mathfrak{G}_{MN} = \left(
                          \begin{array}{cc}
                            (\widehat{F} - B)_{\mu\nu} & - \widehat{D}_\mu \phi^a \\
                            \widehat{D}_\nu \phi^b & \{ \phi^a, \phi^b \}_\theta \\
                          \end{array}
                        \right),
\end{equation}
and
\begin{equation}\label{2open}
  \mathfrak{g}_{\mu\nu} = - \kappa^2 (B g^{-1} B)_{\mu\nu}, \qquad G_s = g_s \sqrt{\det_4 (\kappa B g^{-1})}.
\end{equation}
Substituting the result \eq{app-id1} into Eq. \eq{sdbi-id} leads to an intriguing identity
\begin{eqnarray}\label{strange-id}
    S &=& \frac{1}{2\pi \kappa^2 g_s} \int d^4 y \sqrt{\det_4 \big(\mathcal{G} + \kappa B \big)} \nonumber \\
    &=&  \frac{1}{2\pi \kappa^2 G_s} \int d^4 y \sqrt{\det_{n+4} \big(\mathbb{G} + \kappa \mathfrak{G} \big)}.
\end{eqnarray}

Using the determinant formula for a block matrix
\begin{equation}\label{app-detform}
    \det_{n+4} \left(
           \begin{array}{cc}
             A & B \\
             C & D \\
           \end{array}
         \right) = \det_n D \; \det_4 (A-BD^{-1}C),
\end{equation}
the determinant in Eq. \eq{strange-id} can be written as
\begin{equation}\label{det-prod}
\sqrt{\det_{n+4} \big(\mathbb{G} + \kappa \mathfrak{G} \big)}  = \sqrt{\det_n \mathfrak{q}_{ab}}
\sqrt{\det_4 \big( \mathfrak{g}_{\mu\nu} + \kappa (\widehat{F} - B)_{\mu\nu}
+ \kappa^2 \widehat{D}_\mu \phi^a \mathfrak{q}^{-1}_{ab} \widehat{D}_\nu \phi^b \big)}
\end{equation}
where
\begin{equation}\label{app-q}
  \mathfrak{q}_{ab} = \delta_{ab} +  \kappa \{ \phi^a, \phi^b \}_\theta.
\end{equation}
In the limit $\kappa \to 0$, we get the expansion
\begin{eqnarray}\label{det-expan}
\sqrt{\det_{n+4} \big(\mathbb{G} + \kappa \mathfrak{G} \big)}  &=& \sqrt{\det_4 \mathfrak{g}}
\Big( 1 + \frac{\kappa^2}{4} \mathfrak{g}^{\mu\rho} \mathfrak{g}^{\nu\sigma} (\widehat{F} - B)_{\mu\nu}
(\widehat{F} - B)_{\rho\sigma} + \frac{\kappa^2}{2} \mathfrak{g}^{\mu\nu} \widehat{D}_\mu \phi^a
\widehat{D}_\nu \phi^a \nonumber \\
&& \hspace{2cm} + \frac{\kappa^2}{4} \{ \phi^a, \phi^b \}^2_\theta + \cdots \Big).
\end{eqnarray}
The corresponding NC field theory is obtained by quantizing symplectic gauge fields
and adjoint scalar fields \cite{ncft-rev}.
The quantization in our case is simply defined by the canonical Dirac
quantization of the Poisson algebra $P = \big(C^\infty (M), \{-,-\}_\theta \big)$.\footnote{In general,
the quantization becomes nontrivial if $M$ is a curved manifold described by
the general metric $\mathfrak{g}_{\mu\nu}$. Hence one may take the metric $\mathfrak{g}_{\mu\nu}$
to be flat for simplicity.} The quantization map $\mathcal{Q}: C^\infty (M) \to \mathcal{A}_\theta$
by $f \mapsto \mathcal{Q}(f) \equiv \widehat{f}$ is a $\mathbb{C}$-linear algebra homomorphism
defined by
\begin{equation}\label{q-map}
    f \cdot g \mapsto \widehat{f \star g} = \widehat{f} \cdot \widehat{g}
\end{equation}
and
\begin{equation}\label{star-prod}
    f \star g \equiv \mathcal{Q}^{-1} \big(\mathcal{Q} (f) \cdot \mathcal{Q}(g) \big)
\end{equation}
for $f, g \in C^\infty (M)$ and $\widehat{f}, \widehat{g} \in \mathcal{A}_\theta$.
For example, the quantization replaces the Poisson bracket \eq{app-poisson} by a NC bracket, i.e.,
\begin{equation}\label{app-starbr}
    \{ X^M, X^N \}_\theta  \quad \to \quad -i [\widehat{X}^M, \widehat{X}^N ]_\star.
\end{equation}

After quantization, the nontrivial leading terms in Eq. \eq{det-expan} precisely give rise to
the NC $U(1)$ gauge theory coupled to adjoint scalar fields $\widehat{\phi}^a$ with quartic interactions.
The identity \eq{strange-id}, together with Eq. \eq{det-expan}, clearly verifies that
the coupled system of NC fields $(\widehat{A}_\mu, \widehat{\phi}^a)$ in the commutative limit
can be organized into a four-dimensional Riemannian metric given by \eq{exd-metric}.
We note that the NC fields $(\widehat{A}_\mu, \widehat{\phi}^a)$ produce a universal metric; in other words,
they do not produce their own metrics separately. This is a manifestation of the equivalence principle
in general relativity. We may emphasize that the universal coupling was originated from the Darboux
theorem as the identity \eq{sdbi-id} clearly indicates. Therefore the emergent gravity
realizes the equivalence principle as a noble statement \cite{q-emg} that the local coordinate
transformation to a Darboux frame corresponds to a locally inertial frame in general relativity.

However there is a subtle issue when we consider a charged scalar field $\psi$ in the fundamental representation.
In this case the interaction with NC $U(1)$ gauge fields is given by either
\begin{equation}\label{fl-scalar}
    \widehat{D}_\mu \psi = \partial_\mu \psi - i \widehat{A}_\mu \star \psi
\end{equation}
or
\begin{equation}\label{fr-scalar}
    \widehat{D}_\mu \psi = \partial_\mu \psi + i \psi \star \widehat{A}_\mu.
\end{equation}
It should be noted that the scalar fields in the DBI action \eq{s-dbi} necessarily give rise to NC scalar fields
in the adjoint representation. Thus the coupling in the fundamental representation cannot be deduced
from the Darboux transformation. We do not know yet how to determine gravitational fields emergent from
NC scalar fields in the fundamental representation in the context of emergent gravity.
One way is to adopt the picture in \cite{q-emg} that the scalar field in the fundamental representation
corresponds to a state in a Hilbert space rather than a field. Another way is to take it as a purely
matter part in energy-momentum tensor. Some results in \cite{amb-sw} may be useful for this issue.
We hope to clarify this issue soon.

\section{Spin connections of NC $U(1)$ instantons}

In this appendix, we will show that the (generalized) self-duality equations for NC $U(1)$ instantons
in four and six dimensions can be transformed to geometrical ones for spin connections of an emergent
Riemannian manifold using the Lie algebra homomorphism \eq{homo-f}.
In particular, it is shown that the resulting geometric equations are equivalent to the CY condition
for a Ricci-flat, K\"ahler manifold.

Let us consider the generalized self-duality equations defined by Eq. \eq{nchym3} where
\begin{equation}\label{sde-sc}
    {T_{ab}}^{cd} = \left\{
                      \begin{array}{ll}
                        {\varepsilon_{ab}}^{cd},  & d=4; \\
                        \frac{1}{2}{\varepsilon_{ab}}^{cdef} I_{ef}, \qquad & d=6.
                      \end{array}
                    \right.
\end{equation}
We showed in Sect. 2 that, using the Lie algebra homomorphism \eq{homo-f}, the self-duality
equations \eq{nchym3} can be isomorphically mapped to those of vector fields in
the commutative limit, viz., Eq. \eq{comm-lim}. To be specific, they are given by
\begin{equation}\label{appbps}
    {g_{ab}}^{e} = \pm \frac{1}{2} {T_{ab}}^{cd} {g_{cd}}^{e}
\end{equation}
after using the Lie algebra relation \eq{cvfse}. Let us introduce the Lie bracket for
the frame basis $E_a = E_a^\mu (x) \frac{\partial}{\partial x^\mu} \in \Gamma(T\mathcal{M})$
defined by
\begin{equation}\label{fbst-eq}
    [E_a, E_b] = - {f_{ab}}^c E_c.
\end{equation}
The Lie algebra \eq{fbst-eq} can be rephrased into the structure equations for
the vielbeins $e^a = e^a_\mu (x) d x^\mu \in \Gamma(T^* \mathcal{M})$ given by
\begin{equation}\label{vbst-eq}
    de^a = \frac{1}{2} {f_{bc}}^a e^b \wedge e^c.
\end{equation}
Combining Eq. \eq{vbst-eq} with the torsion free condition, $de^a + {\omega^a}_b \wedge e^b = 0$,
leads to the relation
\begin{eqnarray} \label{spin-stfunc}
f_{abc} &=& \omega_{abc} - \omega_{bac} \nonumber \\
  &=& \omega_{[abc]} - \omega_{cab}
\end{eqnarray}
where we used the symmetrization symbol $\omega_{[abc]} = \omega_{abc} + \omega_{bca} + \omega_{cab}$
for spin connections $\omega_{bc} = \omega_{\mu bc} dx^\mu = \omega_{abc} e^a$.
The structure equations in Eqs. \eq{cvfse} and \eq{fbst-eq} are related to each other
by the relation \eq{vec-emg} \cite{hsy-jhep09,hsy-review}:
\begin{equation}\label{stf-ev}
  {g_{ab}}^{c} = \lambda \big( {f_{ab}}^{c} - E_a \ln \lambda \delta^c_b +
  E_b \ln \lambda \delta^c_a \big).
\end{equation}

Now we want to transform the self-duality equations \eq{appbps} into some equations for spin connections
using the relations \eq{stf-ev} and \eq{spin-stfunc} together with the volume-preserving
condition \eq{vol-v} that is equal to
\begin{equation}\label{app-volpc1}
   \omega_{bab} = f_{bab} = (p+1-d) E_a \ln \lambda
\end{equation}
or equivalently
\begin{equation}\label{app-volpc2}
   g_{bab} = p V_a \ln \lambda.
\end{equation}
First we will check on this approach with the well-established
result in \cite{hsy-jhep09,g-inst} for the four-dimensional case.
Then we will apply it to the six-dimensional case.

In four dimensions, Eq. \eq{appbps} can be written by using Eq. \eq{stf-ev} as
\begin{equation}\label{4-sdv}
    {f_{ab}}^{e} - \phi_{[a} \delta^e_{b]} = \pm \frac{1}{2} {\varepsilon_{ab}}^{cd}
\big( {f_{cd}}^{e} - \phi_{[c} \delta^e_{d]} \big)
\end{equation}
where $\phi_a := E_a \ln \lambda$. Contracting ${\varepsilon_{f}}^{abe}$ on both sides of Eq. \eq{appbps}
leads to the result
\begin{equation}\label{contract-g}
    g_{bab} =  \mp \frac{1}{2} {\varepsilon_{a}}^{bcd} g_{bcd}.
\end{equation}
Using Eqs. \eq{stf-ev} and \eq{app-volpc2}, the above equation can be written as
\begin{equation}\label{cont-gv}
    p \phi_a =  \mp \frac{1}{2} {\varepsilon_{a}}^{bcd} f_{bcd}
\end{equation}
and it can be inverted as
\begin{equation}\label{invert-gv}
  f_{[abc]} =  \pm p {\varepsilon_{abc}}^{d} \phi_d = 2 \omega_{[abc]}.
\end{equation}
Using this result together with the relation \eq{spin-stfunc}, the self-duality equation \eq{4-sdv}
takes the form
\begin{equation}\label{4-sdv1}
    {\omega^{e}}_{ab} + \phi_{[a} \delta^e_{b]} \mp \frac{p}{2} {\varepsilon_{ab}}^{ef} \phi_f
= \pm \frac{1}{2} {\varepsilon_{ab}}^{cd}
\Big( {\omega^{e}}_{cd} + \phi_{[c} \delta^e_{d]} \mp \frac{p}{2} {\varepsilon_{cd}}^{ef} \phi_f  \Big).
\end{equation}
Note that the combination $\phi_{[a} \delta^e_{b]} \mp {\varepsilon_{ab}}^{ef} \phi_f$ automatically
obeys the same type of self-duality equations
\begin{equation}\label{4-sdv2}
    \phi_{[a} \delta^e_{b]} \mp {\varepsilon_{ab}}^{ef} \phi_f
= \pm \frac{1}{2} {\varepsilon_{ab}}^{cd}
\big( \phi_{[c} \delta^e_{d]} \mp {\varepsilon_{cd}}^{ef} \phi_f  \big).
\end{equation}
Subtracting \eq{4-sdv2} from \eq{4-sdv1} gives us the result
\begin{equation}\label{4-sdv3}
    {\omega^{e}}_{ab} \mp \frac{p-2}{2} {\varepsilon_{ab}}^{ef} \phi_f
= \pm \frac{1}{2} {\varepsilon_{ab}}^{cd}
\Big( {\omega^{e}}_{cd} \mp \frac{p-2}{2} {\varepsilon_{cd}}^{ef} \phi_f  \Big).
\end{equation}
Hence the choice $p=2$ adopted in \cite{hsy-jhep09,g-inst} leads to the self-duality equation
for spin connections:
\begin{equation}\label{4-sdv4}
    {\omega^{e}}_{ab} = \pm \frac{1}{2} {\varepsilon_{ab}}^{cd} {\omega^{e}}_{cd}.
\end{equation}

If the spin connection ${\omega^{a}}_{b}$ is (anti-)self-dual, the Riemann curvature tensor ${R^{a}}_{b} =
d{\omega^{a}}_{b} + {\omega^{a}}_{c} \wedge {\omega^{c}}_{b}$ is also (anti-)self-dual, i.e.,
$R_{ab} = \pm \frac{1}{2} {\varepsilon_{ab}}^{cd} R_{cd}$. Conversely, if the curvature tensor
is (anti-)self-dual, the spin connection also becomes (anti-)self-dual up to a gauge choice.
This means that a four-dimensional Riemannian manifold obeying the self-duality equation \eq{4-sdv4}
is a gravitational instanton which is a Ricci-flat, K\"ahler manifold or called a CY 2-fold \cite{egh-prep}.
Thus we have verified the result in \cite{hsy-jhep09,hsy-inst}.

In six dimensions, Eq. \eq{appbps} is similarly written as
\begin{equation}\label{6-sdv}
    {f_{ab}}^{e} - \phi_{[a} \delta^e_{b]} = \pm \frac{1}{2} {T_{ab}}^{cd}
\big( {f_{cd}}^{e} - \phi_{[c} \delta^e_{d]} \big).
\end{equation}
Contracting ${T_{f}}^{abe}$ on both sides of Eq. \eq{appbps} leads to the result
\begin{equation}\label{contract-6g}
    g_{bab} + \frac{1}{2} I^{cd} {g_{cd}}^b I_{ba} =  \mp \frac{1}{2} {T_{a}}^{bcd} g_{bcd}.
\end{equation}
Using Eqs. \eq{stf-ev} and \eq{app-volpc2}, the above equation can be written as
\begin{equation}\label{cont-gv6}
    (p+1) \phi_a + \chi_a =  \mp \frac{1}{2} {T_{a}}^{bcd} f_{bcd}
\end{equation}
where $\chi_a \equiv \frac{1}{2} I^{cd} {f_{cd}}^b I_{ba}$ and it can be inverted as
\begin{equation}\label{invert-gv6}
  f_{[abc]} =  \pm \frac{1}{2} {T_{abc}}^{d} \big( (p+1) \phi_d + \chi_d \big) = 2 \omega_{[abc]}.
\end{equation}
Contracting $\frac{1}{2} I^{ab} I_{eg}$ on both sides of Eq. \eq{6-sdv} and using the identity
$\frac{1}{4} {T_{ab}}^{cd} I_{cd} = I_{ab}$, we get the relation
\begin{equation}\label{6d-two}
    \phi_a = - \chi_a.
\end{equation}
Therefore the relation \eq{invert-gv6} reduces to
\begin{equation}\label{inv-gv6}
  f_{[abc]} =  \pm \frac{p}{2} {T_{abc}}^{d} \phi_d = 2 \omega_{[abc]}.
\end{equation}
Using this result together with the relation \eq{spin-stfunc}, the self-duality equation \eq{6-sdv}
takes the form
\begin{equation}\label{6-sdv1}
    {\omega^{e}}_{ab} + \phi_{[a} \delta^e_{b]} \mp \frac{p}{4} {T_{ab}}^{ef} \phi_f
= \pm \frac{1}{2} {T_{ab}}^{cd}
\Big( {\omega^{e}}_{cd} + \phi_{[c} \delta^e_{d]} \mp \frac{p}{4} {T_{cd}}^{ef} \phi_f  \Big).
\end{equation}
An important step is to find a combination $\phi_{[a} \delta^e_{b]} + \zeta {T_{ab}}^{ef} \phi_f$
which automatically obeys the same type of self-duality equations, i.e.,
\begin{equation}\label{6-sdv2}
    \phi_{[a} \delta^e_{b]} + \zeta {T_{ab}}^{ef} \phi_f
= \pm \frac{1}{2} {T_{ab}}^{cd}
\big( \phi_{[c} \delta^e_{d]} + \zeta {T_{cd}}^{ef} \phi_f  \big).
\end{equation}
It turns out that, unlike the four-dimensional case, such a combination exists only for the $(-)$-equation
with $\zeta = \frac{1}{2}$. It may not be surprising since this case only corresponds to a Ricci-flat,
K\"ahler manifold, i.e., a CY 3-fold. But the sign for the self-duality equation depends on the parity
of the symplectic matrix $I_{ab}$ in Eq. \eq{sde-sc}.
If we flip the orientation of a six-dimensional manifold by choosing a symplectic matrix $I_{ab}$
different from \eq{i-matrix}, instead the $(+)$-equation only may admit such a combination.
In this case a CY 3-fold will be defined by the $(+)$-equations with the different choice of $I_{ab}$.
Subtracting \eq{6-sdv2} from \eq{6-sdv1} with both the lower $(-)$-sign gives us the result
\begin{equation}\label{6-sdv3}
    {\omega^{e}}_{ab} + \frac{p-2}{4} {T_{ab}}^{ef} \phi_f
= - \frac{1}{2} {T_{ab}}^{cd}
\Big( {\omega^{e}}_{cd} + \frac{p-2}{4} {T_{cd}}^{ef} \phi_f  \Big).
\end{equation}
Therefore, as in the four-dimensional case, the choice $p = 2$ leads to the desired self-duality equation
for spin connections:
\begin{equation}\label{6-sdspin}
    {\omega^{e}}_{ab} = - \frac{1}{2} {T_{ab}}^{cd} {\omega^{e}}_{cd}.
\end{equation}

\end{document}